\documentstyle[multicol,aps,prb,psfig]{revtex}

\begin{document}


\title{Influence of disorder on the ferromagnetism in 
diluted magnetic semiconductors.}

\author{A. L. Chudnovskiy, D. Pfannkuche \\
I Institut f\"ur Theoretische Physik,  Universit\"at Hamburg,  \\
Jungiusstr. 9, D-20355 Hamburg, Germany}

\maketitle

\begin{abstract}
Influence of disorder in the concentration of magnetic Mn$^{2+}$ ions on the
ferromagnetic phase transition in diluted (III,Mn)V
semiconductors is investigated analytically.
The regime of small disorder is addressed, and the enhancement of
the critical temperature by disorder is found both in the mean field approximation and
from the analysis of the zero temperature spin stiffness. Due to disorder, the
spin wave fluctuations around the ferromagnetically ordered state acquire a finite mass.
At large charge carrier band width, the spin wave mass squared becomes negative, signaling
the breakdown of the ferromagnetic ground state.
\end{abstract}

\section{Introduction}

The prospects of application in spin electronic devices inspired much experimental \cite{Ohno} 
and
theoretical \cite{MacDonald,Bhatt,Tc,Litvinov,Schliemann}
activity in studying the itinerant  ferromagnetism in (III,Mn)V diluted magnetic semiconductors.
The theoretical efforts led to much understanding of the basic features of that material system,
providing a mean-field description and reasonable estimates of the ferromagnetic phase transition
temperature from the spin-wave  stiffness \cite{MacDonald,Tc} .

There are two important issues, however, that have not been completely understood so far,
namely, the role of disorder and interactions  between
the mobile charge carriers (holes) in the ferromagnetic phase transition.
In this paper, we address the issue of disorder in the concentration of magnetic Mn-dopants.
The role of disorder is two-fold.
On one hand, the variation in the Mn-concentration creates  regions with local
concentration of Mn-spins higher than the average. These regions
tend to be magnetically polarized by itinerant carriers at higher temperatures than the
phase transition
temperature for the homogeneous system.   The latter should lead to higher transition
temperatures in the disordered system. On the other hand,
the disorder leads to localization of the mobile charge carriers, thus hampering the onset of
long range magnetic correlations,
which would lower the transition  temperature in the disordered system. Presumably, the first effect prevails,
if the localization of holes is weak,  whereas in the case of  strongly localized carriers (strong disorder)
the second effect takes over.
The enhancement of the transition temperature in disordered systems has been reported in recent theoretical investigations
by R. N. Bhatt and M. Berciu \cite{Bhatt}, and J. Schliemann {\it et al} \cite{Tc}.
The results of this paper qualitatively agree with \cite{Bhatt,Tc}. However, the
region of strong localization of the charge carries, where the suppression of the critical temperature presumably takes place
is not addressed here.

\section{Theoretical model and approximations}

The microscopic description of the itinerant ferromagnetism in (III,Mn)V semiconductors is given by the Kondo
lattice Hamiltonian. In contrast to the Kondo system, in the system under consideration the
concentration of the charge carriers (mobile holes in the valence band) $p$ is much lower
than the concentration of the Mn-spins $N_{Mn}$.
The latter rules out the Kondo effect. Moreover, the $pd$ exchange coupling $J$ is larger than the Fermi energy
\cite{Ohno,MacDonald}.
As a result, in the ferromagnetic state, the charge carriers (holes) are completely spin
polarized.

A hole magnetically polarizes the Mn spins in its vicinity, thus forming a magnetic polaron.
Close to the ferromagnetic phase transition there is a large correlation
length in the Mn spin system, and the dynamics of mobile holes
is much faster than the dynamics of Mn spins. In this case, the magnetic polarization cloud
created by a hole stays for some relaxation time $\tau_{Mn}$ after the hole has
left the region.
The linear size of the polaron cloud can be estimated
as $R\sim v\tau_{Mn}$, where $v$ is the velocity
of the hole (the motion of the hole inside the polarized region is assumed to be ballistic).
The condition of low hole concentration,
$p\ll N_{Mn}$, \cite{Ohno,MacDonald}
implies that there is no more than one mobile hole polarizing the cloud at a
given moment of time.

Therefore, the physical picture close to the phase transition point is that of
slowly relaxing Mn-clusters (or polaronic clouds) polarized by the spins of mobile holes.
The hopping of holes establishes long range correlations between the clusters.
Due to the randomness in the Mn concentration,
the number of Mn-spins in each cluster is randomly distributed around some average value.
This picture is conveyed in the mathematical model below.

The model consists of mobile holes moving on a lattice with sites $i$.
The holes are represented by the
fermionic operators $\hat{a}^+_{i\sigma}, \hat{a}_{i\sigma}$. The other ingredient
is a system of Mn-spins $5/2$ distributed randomly over a lattice with sites $j$.
Since the concentration of Mn spins is larger then the concentration of holes
\cite{Ohno,MacDonald}, there are many Mn-sites within a unit cell of the
hole lattice.
A hole polarizes Mn-spins in its vicinity. Because of the randomness in the
Mn distribution, the number of polarized spins around the site $i$ is random.

The Hamiltonian of the model writes
\begin{equation}
\hat{H}=-t\sum_{<i,i'>}(\hat{a}^+_{i\sigma}\hat{a}_{i'\sigma}+h.c.) +
J\sum_i\sum_{j=1}^{N_i}{\bf\hat{s}}_i{\bf\hat{S}}_{ji}.
\label{H}
\end{equation}
The first term describes the nearest neighbor hopping of holes. In this paper, a simplified
one-band hole dispersion is used instead of the six-band model relevant for (Ga,Mn)As.
\cite{MacDonald}  Whereas the one-band model is insufficient to give a quantitative
prediction for the critical temperature, it reproduces the basic qualitative features of
the hole system, including the effects of disorder \cite{Bhatt,Schliemann}.   In the interaction term,
${\bf\hat{s}}_i=\hat{a}^+_{i\alpha}{\bf\sigma}_{\alpha\beta}\hat{a}_{i\beta}$ describes the
spin of a hole on
site $i$, and the operators ${\bf\hat{S}}_{ji}$ describe the Mn-spins in the vicinity of
site $i$.   The interaction term is written under the assumption of a rectangular hole wave
function, so that the constant $J$ does not depend on the distance between
the hole and the Mn-spin within the interaction range.

There is a total of $N_i$ Mn-spins polarized by the hole at site $i$.
The number $N_i$ is taken to be random gaussian distributed with mean $N_p$ and variation
$\Delta/\sqrt{N_p}$
 \begin{equation}
 P(N_i)\propto \exp\left[-\frac{N_p(N_i-N_p)^2}{2\Delta^2}\right].
 \label{PN}
 \end{equation}
In the experimental system the average $N_p$ is usually small \cite{Ohno},
$N_p\sim 3$. Therefore, the Poissonian distribution is actually more
appropriate to describe the disorder in $N_i$. We take the Gaussian distribution for the sake
of technical convenience. The case of the Poissonian distribution of $N_i$ is left for future
investigations.
To restrict the influence of the unphysical region of negative $N_i$, the
condition $\Delta\ll N_p$ is adopted in numerical evaluations below (see Section \ref{secMF}).

To deal with disorder, we employ the replica-trick.  In the replicated partition function for the
Hamiltonian (\ref{H}), the interaction term is represented by
$\exp[-J\sum_i \sum_{j=1}^{N_i}\sum_{a=1}^n\int_0^\beta d\tau{\bf s}^{a\tau}_i
{\bf S}^{a\tau}_{ji}]$,
where $a$ denotes the
replica-index, and $\tau$ denotes the imaginary time.  We decouple the interaction term by the
Hubbard--Stratonovich
transformation, introducing decoupling fields for the magnetization of Mn-spins and
hole-spins. After
the decoupling, the interaction part of the partition function assumes the form
\begin{eqnarray}
\nonumber &&
\exp[-J\sum_i \sum_{j=1}^{N_i}\sum_{a=1}^n\int_0^\beta d\tau{\bf s}^{a\tau}_i
{\bf S}^{a\tau}_{ji}]
\propto \int\prod_{<i,j>}D[{\bf X}_{ij}^{a\tau}]D[{\bf Y}_{ij}^{a\tau}]\\
&&
\exp\left[-\sum_a\int_0^\beta d\tau \sum_{<i,j>}\frac{{\bf X}_{ij}^{a\tau}{\bf Y}_{ij}^{a\tau}}{J}
+\sum_a\int_0^\beta d\tau \sum_{<i,j>}\left\{{\bf X}_{ij}^{a\tau}{\bf S}_{ij}^{a\tau}
-{\bf Y}_{ij}^{a\tau}{\bf s}_{i}^{a\tau}\right\}\right].
\label{dec-int}
\end{eqnarray}
The term $\sum_{<i,j>}{\bf{X}}_{ij}^{a\tau}{\bf{S}}_{ij}^{a\tau}$ describes the Mn-spins in the
cluster around
site $i$ under the influence of the effective field created by mobile holes.
${\bf{X}}_{ij}^{a\tau}$ is the effective field created by a hole at site $i$ and acting on the
Mn-spin on site $j$. For a
rectangular form of the hole wave function,
we replace ${\bf X}_{ij}^{a\tau}={\bf X}_{i}^{a\tau} \  \forall j=\overline{1,N_i}$.
In what follows we use the Holstein--Primakoff representation for the total spin of the cluster
\cite{HPrep},
${\bf\Xi}_i\equiv\sum_{j=1}^{N_i}{\bf S}_{ij}$,
\begin{eqnarray}
\Xi_i^z&=&N_i S-\hat{b}^+_i\hat{b}_i , \\
\Xi_i^+&=&(\sqrt{2N_iS-\hat{b}^+_i\hat{b}_ i})\hat{b}_i, \\
\Xi_i^-&=&\hat{b}_i^+(\sqrt{2N_iS-\hat{b}^+_i\hat{b}_ i}), \\
\nonumber
\end{eqnarray}
where $\hat{b}_i^+, \hat{b}_i$ are  bosonic creation and annihilation operators, describing the
excitations  around the
ordered state of the cluster. There is a constraint on the number of bosons in each cluster
$\langle \hat{b}_i^+\hat{b}_i\rangle\leq 2N_i S$, $S=5/2$. The further calculations are carried
out in the quadratic approximation in the bosonic fields  $\hat{b}_i^+, \hat{b}_i$ ,
so we replace in the square root
$\sqrt{2N_iS-\hat{b}^+_i\hat{b}_ i}\approx\sqrt{2N_iS}$.

Finally, under the assumption of the rectangular form of the hole's wave function,
we approximate the effective field exerted
by the Mn-cluster on the hole at the site $i$, $\sum_{j=1}^{N_i}{\bf Y}_{ij}^{a\tau}
{\bf s}_{i}^{a\tau}\approx N_i {\bf Y}_i^{a\tau}{\bf s}_{i}^{a\tau}$.
Under the approximations above, the replicated partition function of the model (\ref{H})
can be written as
\begin{eqnarray}
 \nonumber &&
 Z^n\approx\int D[\bar{a}_{i\sigma}^{a\tau}] D[a_{i\sigma}^{a\tau}] D[\bar{b}_{i\sigma}^{a\tau}]
 D[b_{i\sigma}^{a\tau}]
 D[{\bf X}_i^{a\tau}]D[{\bf Y}_i^{a\tau}]  \exp\left[\sum_{a=1}^n\sum_{i}\int_0^{\beta}d\tau
 \left\{\sum_{\sigma=\uparrow,\downarrow}\bar{a}_{i\sigma}^{a\tau}(-\partial_{\tau}+\mu)
 a_{i\sigma}^{a\tau}\right.\right.\\
\nonumber &&
\left.+\bar{b}_i^{a\tau}(-\partial_{\tau}-X_{zi}^{a\tau})b_i^{a\tau}\right\}
+t\sum_{<i,i'>}\sum_{a=1}^n\sum_{\sigma}(\bar{a}^{a\tau}_{i\sigma}a^{a\tau}_{i'\sigma}+
 \bar{a}^{a\tau}_{i'\sigma}a^{a\tau}_{i\sigma}) \\
 \nonumber &&
 -\sum_i N_i\sum_a\int_0^{\beta}d\tau\left\{Y_ {zi}^{a\tau}({\bf\bar{a}}_i^{a\tau}
{\bf\sigma}^z{\bf a}_i^{a\tau})+
SX_{zi}^{a\tau}-(X_{zi}^{a\tau}Y_{zi}^{a\tau})/J\right\} \\
&&\nonumber
+\sum_i\sum_{a=1}^n\int_0^{\beta}d\tau\left\{\sqrt{2SN_i}(X_{i+}^{a\tau}\bar{b}_i^{a\tau}+
X_{i-}^{a\tau}b_i^{a\tau})-N_i\left(Y_{i+}^{a\tau}({\bf\bar{a}}_i^{a\tau}
{\bf\sigma}^-{\bf a}_i^{a\tau})+
Y_{i-}^{a\tau}({\bf\bar{a}}_i^{a\tau}{\bf\sigma}^+{\bf a}_i^{a\tau})\right) \right.\\
&& \left.\left.
-N_i (X_{i+}^{a\tau}Y_{i-}^{a\tau}+X_{i-}^{a\tau}Y_{i+}^{a\tau})/J\right\}\right].
\label{Zn}
\end{eqnarray}
Here $\bar{a}_{i\sigma}^{a\tau},  a_{i\sigma}^{a\tau}$ denote the Grassman fields associated 
with the mobile holes, and
 $\bar{b}_{i}^{a\tau},  b_{i}^{a\tau}$ denote the complex variables associated with the bosonic
 operators
 $\hat{b}_i^+, \hat{b}_i$. We introduced the spinors ${\bf a}_{i}^{a\tau}, {\bf\bar{a}}_{i}^{a\tau}$,
 which are formed as follows,
 ${\bf a}_{i}^{a\tau}\equiv (a_{i\uparrow}^{a\tau},  a_{i\downarrow}^{a\tau})^T$,
 ${\bf\bar{a}}_{i}^{a\tau}=({\bf a}_{i}^{a\tau})^+$.

 The partition function (\ref{Zn}) is analyzed on the mean field level. The mean field
 approximations include neglecting
 the transverse fluctuations of spins (only $z$-components of fields are considered), and a
 static and spatially homogeneous
 ansatz for the fields $X_{iz}^{a\tau}\approx X^{a}$,  $Y_{iz}^{a\tau}\approx Y^{a}$.
 The averaging over the disorder with the distribution (\ref{PN})
results in a term in the effective free energy that mixes different replicas and looks as
 \begin{equation}
 -\frac{\Delta^2}{2N_p}\sum_i\sum_{a,b=1}^n Y^aY^b\int_0^{\beta}d\tau d\tau'
 ({\bf\bar{a}}_i^{a\tau}{\bf\sigma}^z{\bf a}_i^{a\tau})({\bf\bar{a}}_i^{b\tau'}{\bf\sigma}^z{\bf a}_i^{b\tau'}).
 \label{abmix}
 \end{equation}
The structure of term (\ref{abmix}) is typical for the replica treatment of the Anderson
localization problem
\cite{Wegner}.  Under the mean field approximations made above, the strength of the disorder
in the effective localization
problem is proportional to the mean field value of the
squared magnetization in the Mn-clusters $Y^aY^b$. Therefore,
expression (\ref{abmix}) accounts for the localization of holes by the random potential from the
frozen Mn-clusters.
The  treatment of (\ref{abmix}) repeats the nonlinear sigma-model approach for the
Anderson localization
\cite{Wegner}.
The term is decoupled in the following fashion
\begin{eqnarray}
\nonumber &&
\exp\left[-\frac{\Delta^2}{2N_p} \sum_i\sum_{a,b=1}^n Y^aY^b\int_0^{\beta}d\tau d\tau'
 ({\bf\bar{a}}_i^{a\tau}{\bf\sigma}^z{\bf a}_i^{a\tau})({\bf\bar{a}}_i^{b\tau'}{\bf\sigma}^z{\bf a}_i^{b\tau'})
 \right]  \propto \\
 &&
 \nonumber
 \int d{\bf Q}_i^{a\tau,b\tau'}d{\bf Q}_i^{b\tau',a\tau} \exp\left[-\frac{N_p}{2\Delta^2}
 \sum_i\sum_{a,b=1}^n\left\{ Y^aY^b\int_0^{\beta}d\tau d\tau' [{\bf Q}_i^{a\tau,b\tau'}(\sigma^z\otimes\sigma^z)
 {\bf Q}_i^{b\tau',a\tau}\right.\right.\\
  && \left.\left.
  +i\frac{Y^aY^b}{2}\left\{{\bf Q}_i^{a\tau,b\tau'}(\sigma^z\otimes\sigma^z){\bf\bar{a}}_i^{b\tau'}
 {\bf a}_i^{a\tau}+{\bf\bar{a}}_i^{a\tau}{\bf a}_i^{b\tau'}(\sigma^z\otimes\sigma^z)
 {\bf Q}_i^{b\tau',a\tau} \right\}]
 -\frac{\beta^2}{2}\log(Y^aY^b)\right\}\right],
\label{Qdec}
\end{eqnarray}
where ${\bf Q}_i^{a\tau,b\tau'}$ is a matrix in the spin space with elements $(Q_i^{a\tau,b\tau'})_{\sigma\sigma'}$.
Here we concentrate only on the mean field solution for the matrix ${\bf Q}$.  After
Fourier transformation  with respect to
imaginary times $\tau, \tau'$, the mean field ansatz reads
\begin{equation}
 {\bf Q}_i^{a\omega,b\omega'}={\bf\Lambda}^a_\omega  sign(\omega)\delta_{\omega\omega'}\delta_{ab}.
 \label{Qmf}
 \end{equation}
Neglecting  the transverse spin fluctuations, we assume a diagonal structure of the matrix
${\bf\Lambda}_\omega$ in the spin sector
\begin{equation}
{\bf\Lambda}^a_\omega=\left(
\begin{array}{cc}
\Lambda_{\uparrow\uparrow}^{a,\omega} & 0\\
0 & \Lambda_{\downarrow\downarrow}^{a,\omega}
\end{array}
\right).
\label{Lambda}
\end{equation}
Finally, the mean field partition function acquires the form
 \begin{eqnarray}
 \nonumber &&
\langle Z^n\rangle_{MF}=\int D[\bar{a}_{i\sigma}^{a\tau}] D[a_{i\sigma}^{a\tau}]
D[\bar{b}_{i\sigma}^{a\tau}]D[b_{i\sigma}^{a\tau}]
\exp\left[N\sum_{a=1}^n\sum_{\omega_n}\int d^d k
 \left\{\bar{a}_{k\uparrow}^{a\omega}[i\omega_n+\mu-\epsilon_k-N_pY^a \right.\right. \\
 \nonumber &&
 +iY^a\Lambda_{\uparrow\uparrow}^{a\omega}
 sign(\omega_n)+\beta\frac{\Delta^2}{N_p}Y^a\sum_{b=1}^n(Y^bX^b/J-SX^b)]a_{k\uparrow}^{a\omega} \\
 \nonumber && \left.
 +\bar{a}_{k\downarrow}^{a\omega}[i\omega_n+\mu-\epsilon_k+N_pY^a+
 iY^a\Lambda_{\downarrow\downarrow}^{a\omega}
 sign(\omega_n)-\beta\frac{\Delta^2}{N_p}Y^a\sum_{b=1}^n(Y^bX^b/J-SX^b)]
 a_{k\downarrow}^{a\omega}\right\} \\
 \nonumber &&
+ \sum_{a=1}^n\sum_{\omega_n}\sum_{i=1}^N\left\{
\bar{b}_i^{a\omega}(-i\omega_n-X^{a})b_i^{a\omega}\right\}-\frac{NN_p}{2\Delta^2}\sum_{a=1}^n(Y^a)^2
\sum^F_\omega\left\{(\Lambda_{\uparrow\uparrow}^\omega)^2+
(\Lambda_{\downarrow\downarrow}^\omega)^2\right\} \\
&& \nonumber
+\beta NN_p\sum_{a=1}^n\left\{SX^{a}-X^{a}Y^{a}/J\right\}
+\frac{N\beta^2\Delta^2}{2N_p}\sum_{a,b=1}^n\left\{S^2X^aX^b+X^aX^bY^aY^b/J^2 \right. \\
&& \left.\left.
-2SX^aX^bY^b/J\right\}
\right].
\label{Zmf}
\end{eqnarray}
Here the sum over the frequencies $\omega_n$ denote the sum over the
fermionic Matsubara frequencies for the fields $\bar{a}, a$, and the sum over the 
bosonic Matsubara frequencies for the fields $\bar{b}, b$.  
$N$ denotes the total number of sites $i$ in the lattice.
The exponent in (\ref{Zmf}) is quadratic in  the bosonic ($\bar{b}, b$) and the fermionic ($\bar{a}, a$) variables,
hence the bosons and fermions can be intergrated out. When integrating out the bosons, the
restriction on the bosonic occupation number $n_b\leq 2N_pS$ is explicitely taken into account.

\section{Mean field equations and their solutions}
\label{secMF}

We adopt the replica symmetric mean field ansatz for the fields $Y^a=Y$,
$X^a=X$,   $\Lambda^{a\omega}_{\sigma\sigma}=
\Lambda^\omega_{\sigma\sigma} \ \forall a=\overline{1,n}$. In the replica symmetic approximation,
the terms in
(\ref{Zmf}) involving the sum over different replicas give no contribution to the mean field
equations in the replica limit $n\rightarrow 0$.

Further,  we change from the integral over wave vectors to an integral over the energies
$\int d^d k ... = \int\rho_d(\epsilon) d\epsilon$, where $\rho_d(\epsilon)$ denotes the
d-dimensional density of states (DoS).
Under the above approximations, the following mean field equations have been obtained
\begin{equation}
Y=\frac{J}{N_p}\frac{\sum_{l=0}^{(5N_p-1)/2}\left(\frac{5}{2}N_p-l\right)
\sinh\left[\beta\left(\frac{5}{2}N_p-l\right)X\right]}{\sum_{l=0}^{(5N_p-1)/2}
\cosh\left[\beta\left(\frac{5}{2}N_p-l\right)X\right]}.
\label{Ymf}
\end{equation}
Here $Y$ is the average magnetization of a single Mn-spin in units of $J$.
Mn-spins are magnetically polarized by the effective field
of the holes $X$, which is reflected by the RHS of Eq. (\ref{Ymf}).
The average magnetization of a hole $X$ (in units of $J$) is given by the equation
\begin{eqnarray}
\nonumber &&
X=\frac{J}{\Delta^2}YT\sum^F_{\omega_n}\left\{(\Lambda_{\uparrow\uparrow}^{\omega})^2+
(\Lambda_{\downarrow\downarrow}^{\omega})^2\right\}\\
\nonumber &&
+JT\sum_{\omega_n}^F\int_{-W}^{W}\rho(\epsilon)d\epsilon
\left\{\frac{1}{i\omega_n+\mu-\epsilon+N_pY+
iY^2\Lambda_{\downarrow\downarrow}^{\omega}sign(\omega_n)}\right. \\
&& \left.
-\frac{1}{i\omega_n+\mu-\epsilon-N_pY+
iY^2\Lambda_{\uparrow\uparrow}^{\omega}sign(\omega_n)}\right\}.
\label{Xmf}
\end{eqnarray}
Eq. (\ref{Xmf}) describes the magnetic polarization of itinerant carriers (holes)
in the effective field $N_pY$ created by Mn-spins.
Disorder enters Eq. (\ref{Xmf}) through the disorder strength $\Delta$ and
the fields $\Lambda_{\sigma\sigma}^{\omega}$ which, on the mean field level, 
are proportional to the inverse mean free time for holes.
At zero temperature, the inverse mean free time for holes with a spin projection
$\sigma$ is given by
$1/\tau_{\sigma}=2Y^2\Lambda_{\sigma\sigma}^0$. Since only the scattering by
frozen magnetic moments is taken into account on the mean field level, the influence of
disorder disappears completely in the paramagnetic phase.  In the paramagnetic phase,
the scattering by dynamical magnetic fluctuations in Mn spin clusters affects the hole
motion. The account for that effect is beyond the static mean field approximation adopted
here. The analysis of that and other fluctuation effects beyond the mean field approximation
is left for future investigations.

Equations for the mean field values of
$\Lambda_{\sigma\sigma}^{\omega}$ can be written as
\begin{equation}
\Lambda_{\sigma\sigma}^{\omega}=\frac{\Delta^2}{N_p}\int_{-W}^W
\rho(\epsilon)d\epsilon\frac{i \ sign(\omega_n)}{i\omega_n+\mu-\epsilon\mp N_pY
+iY^2\Lambda_{\sigma\sigma}^{\omega} \ sign(\omega_n)}.
\label{Lass}
\end{equation}
The upper sign in the denominator corresponds to $\sigma=\uparrow$ and the lower sign
corresponds to $\sigma=\downarrow$.
For $d=2,3$,  I replace  the density
of states $\rho_d(\epsilon)$ in (\ref{Xmf}), (\ref{Lass}) by a constant average
density of states $\rho_d(\epsilon)=\rho$. The example of a spherically
symmetric band $\epsilon({\bf k})=-W\cos(|{\bf k}|)$ shows that for
large $W$, the contribution of the
singularity of the DoS near the band edge $\epsilon=W$ is small in both two and 
three dimensions,
hence the above replacement is justified.  Upon that replacement, the integrals
over the energy $\epsilon$ in Eqs. (\ref{Xmf}), (\ref{Lass}) can be performed explicitely,
and the equations for $X$ and $\Lambda_{\sigma\sigma}$ assume the form
\begin{eqnarray}
\nonumber &&
X=J\rho T\sum^F_{\omega_n>0}\log\left|\frac{[(W-\mu+N_pY)^2+(\omega_n+
Y^2\Lambda_{\uparrow\uparrow}^\omega)^2]
[(W+\mu+N_pY)^2+(\omega_n+
Y^2\Lambda_{\downarrow\downarrow}^\omega)^2]}{[(W+\mu-N_pY)^2+(\omega_n+
Y^2\Lambda_{\uparrow\uparrow}^\omega)^2]
[(W-\mu-N_pY)^2+(\omega_n+Y^2\Lambda_{\downarrow\downarrow}^\omega)^2]}\right| \\
&&
+\frac{J}{\Delta^2}YT\sum^F_{\omega_n}\left\{(\Lambda_{\uparrow\uparrow}^\omega)^2+
(\Lambda_{\downarrow\downarrow}^\omega)^2\right\}.
\label{Xmf1}
\end{eqnarray}
\begin{eqnarray}
\nonumber &&
\Lambda_{\uparrow\uparrow}^\omega=\rho\frac{\Delta^2}{N_p}\left[\arccos
\left(\frac{N_pY-W-\mu}{|N_pY-W-\mu+i(\omega_n+Y^2\Lambda_{\uparrow\uparrow}^\omega)|}\right)
\right. \\
&& \left.
-\arccos\left(\frac{N_pY+W-\mu}{|N_pY+W-\mu+i(\omega_n+Y^2\Lambda_{\uparrow\uparrow}^\omega)|}
\right)\right],  \ \omega_n>0,
\label{Laup}
\end{eqnarray}
\begin{eqnarray}
\nonumber &&
\Lambda_{\downarrow\downarrow}^\omega=\rho\frac{\Delta^2}{N_p}\left[
\arccos\left(\frac{N_pY-W+\mu}{|N_pY-W+\mu+i(\omega_n+Y^2\Lambda_{\downarrow\downarrow}^\omega)|}
\right)\right. \\
&& \left.
-\arccos
\left(\frac{N_pY+W+\mu}{|N_pY+W+\mu+i(\omega_n+Y^2\Lambda_{\downarrow\downarrow}^\omega)|}
\right)\right],   \
\omega_n>0.
\label{Ladown}
\end{eqnarray}

We analyze the mean field phase diagram by solving numerically mean field equations
(\ref{Ymf}), (\ref{Xmf1})--(\ref{Ladown}), taking the parameters close to
those of (Ga,Mn)As.
The exchange constant for (Ga,Mn)As equals $J_0\approx55$ meV nm$^3$.
The constant $J_0$ is adopted in the model with a $\delta$-like interaction
between a hole and Mn-spins of the form \cite{MacDonald}
\begin{equation}
H_{int}=-J_0\sum_{i,I}{\bf S}_I{\bf s}_i\delta({\bf r}_i-{\bf R}_I).
\label{J0}
\end{equation}
Comparison of (\ref{J0}) with the interaction term in (\ref{H}) for the 
homogeneous case $N_i=N_p$ gives the relation $J=J_0/N_p$.
The number $N_p$ can be evaluated as $N_p\sim N_{Mn}/p$.
For $N_{Mn}=1.0$ nm$^{-3}$, $p=0.35$ nm$^{-3}$ we infer $N_p\sim (N_{Mn}/p)\approx 3$.
For the simplified band structure adopted here, the average density of states
$\rho$ and the width of the band $W$ cannot be directly obtained from experimental data.
The chemical potential is fixed by the condition of constant concentration
of holes $p$, or constant filling factor $\nu$ of the charge carrier lattice $i$ of the
model (\ref{H}).

\begin{figure}
\psfig{file=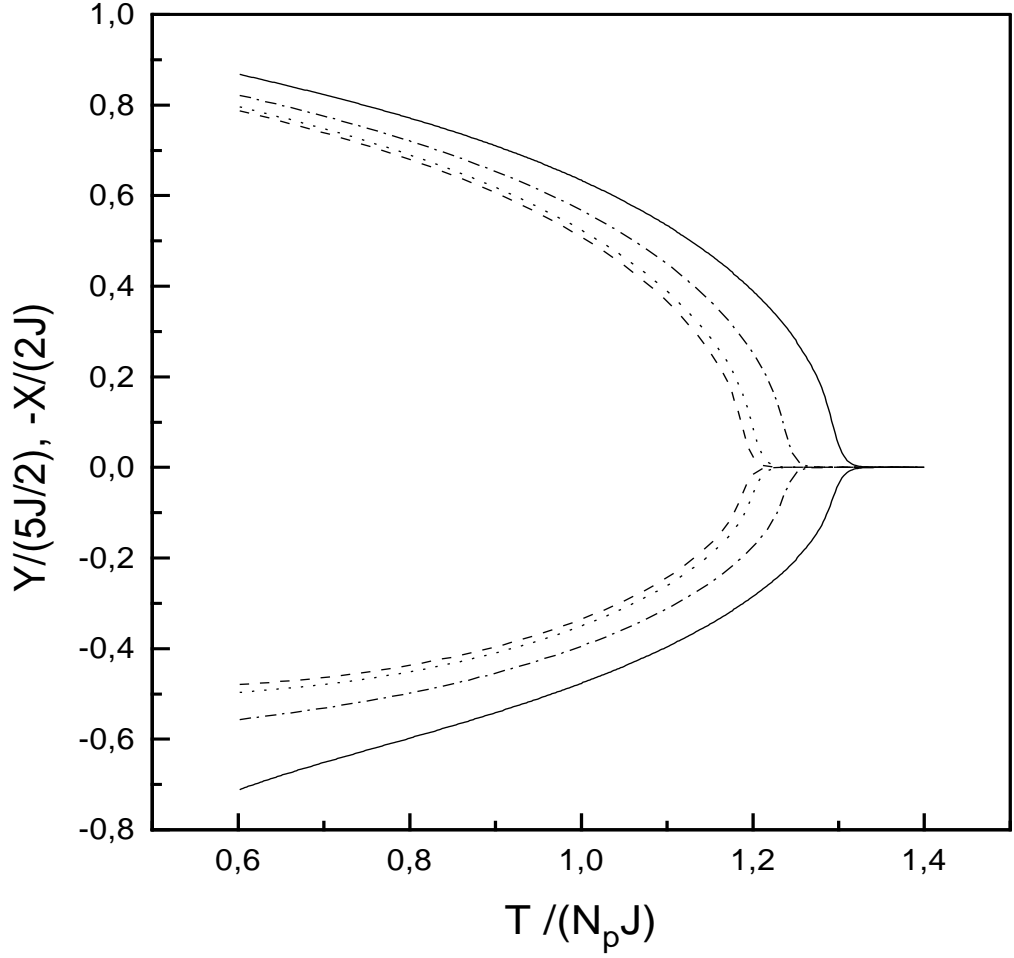,width=15cm,height=20cm,angle=0}
\vskip -3cm
\caption{Relative magnetization of Mn-spins $Y/(JS)$  ($\geq 0$) and holes $-X/(2J)$
($\leq 0$) versus temperature at different disorder strengths. The critical temperature
increases with the disorder strength $\Delta$. In order of increasing $T_c$:
$\Delta=0.1; 1.0; 2.0; 3.0$. Other parameters: $J=10$ meV nm$^3$,
$N_p=5$, $W=100$ meV, $\rho=0.01$ meV$^{-1}$nm$^{-3}$,
$\nu=0.25$.} \label{figXYT}
\end{figure}

The mean field equations (\ref{Ymf}), (\ref{Xmf1}) -- (\ref{Ladown})
were solved numerically for $J=10$ meV, and  different values of
$N_p$, $\rho$, $W$,  $\nu$, and $\Delta$. In all cases we found that the critical
temperature
increases with disorder strength $\Delta$.  The temperature dependence of
the magnetizations of holes $X$ and Mn-spins $Y$ for the set of parameters
$J=10$ meV nm$^3$, $N_p=5$, $W=100$ meV, $\rho=0.01$ meV$^{-1}$ nm$^{-3}$,
$\nu=0.25$ and different values of disorder strength $\Delta$ is shown in Fig. \ref{figXYT}.

\section{Spin waves in a disordered ferromagnet}

There are massless Goldstone modes in the ordered phase of a clean system --
the spin waves.
The zero temperature spin stiffness has been used in Ref. \cite{Tc} to give
an estimate for the critical temperature, which is much closer to experimental data than the
mean field result in the regime of small concentration of charge carriers and large
exchange coupling $J$ (as compared to the Fermi energy).

The spin wave excitations are described by bosonic fields $\bar{b}, b$.
In this section, we calculate the zero temperature dispersion of the spin-wave excitations
around the saddle point (\ref{Ymf}) -- (\ref{Ladown}) described above, and use the result to
analyze qualitatively
the influence of disorder on the critical temperature calculated from the spin wave stiffness.
Since the role of disorder increases at low temperatures, the calculation of
the zero temperature spin stiffness using the mean field description of the disorder effects
(the fluctuations of the field $\bf Q$ are not taken into account) is perturbative in the
disorder strength $\Delta$. It can be used only to describe the qualitative behavior of $T_c$.
The zero temperature spin wave dispersion is given by
\begin{equation}
\Omega=Dq^2+m_{SW}^2.
\label{disp-b}
\end{equation}
The spin waves acquire a mass $m_{SW}$.
Both $D$ and $\lambda$ also have imaginary parts that describe the attenuation of
spin waves in a disordered system. At zero temperature, the approximate expression for the
spin stiffness reads
\begin{equation}
D\approx \frac{2\pi\rho v_F^2}{3N_pS}\left(1+4\frac{\Delta^2}{N_p^5}\right),
\label{Sstiff}
\end{equation}
where we neglected the imaginary part and assumed $\pi\rho\ll N_pJ$, which is consistent
with the regime of low concentration of holes. $v_F$ denotes the Fermi velocity.
One can see from expression
(\ref{Sstiff}) that the spin stiffness grows with disorder strength $\Delta$ at small
disorder. This in turn results in an increase of the critical temperature. The critical
temperature can be estimated according to formula \cite{Tc}
\begin{equation}
T_c\approx D k^2_D
\label{TcSw}
\end{equation}
with the Debye wave vector $k_D=(6\pi^2N_p)^{1/3}$.

\begin{figure}
\psfig{file=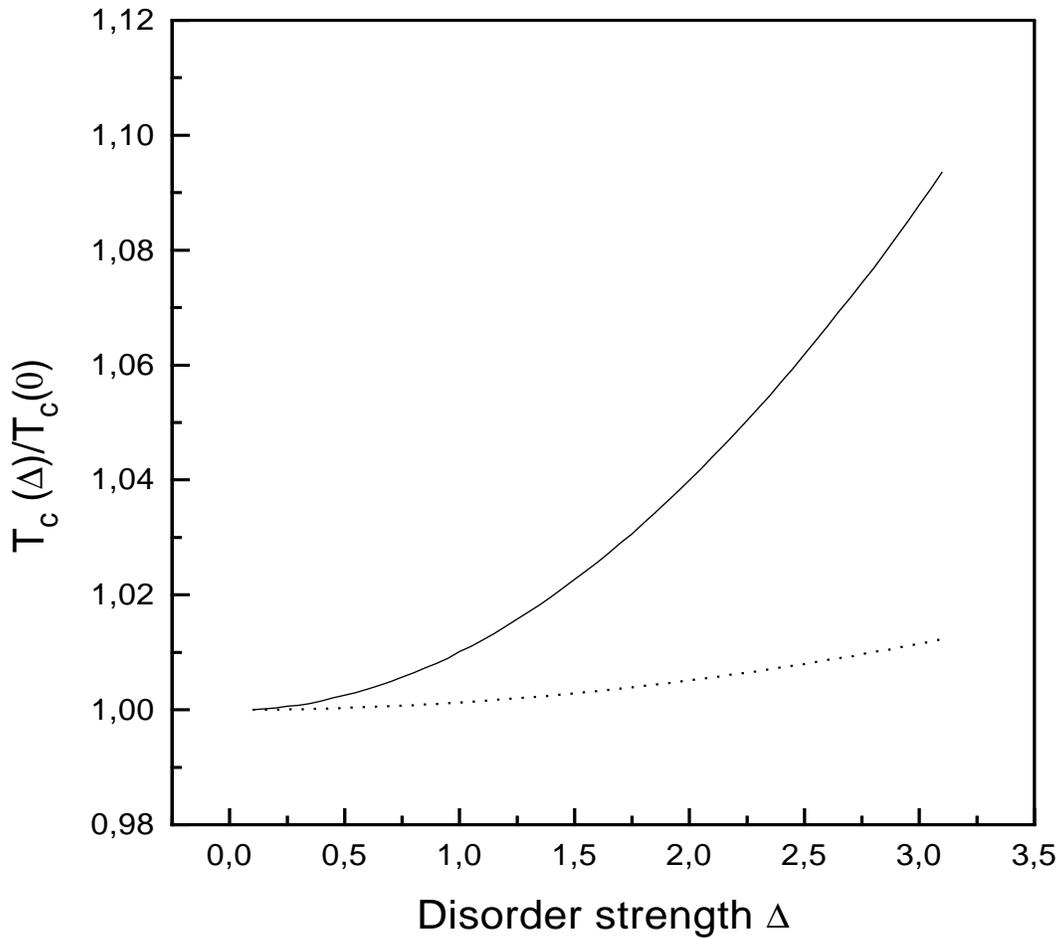,width=15cm,height=20cm,angle=0}
\vskip -3cm
\caption{Relative change of the critical temperature with the strength of disorder 
$T_c(\Delta)/T_c(\Delta=0)$  as calculated from
the mean field equations  (solid line), and from  the 
zero-T spin wave stiffness (dots).
$J=10$ meV nm$^3$, $N_p=5$, $W=100$ meV, $\rho=0.01$meV$^{-1}$nm$^{-3}$, $\nu=0.25$.}
\label{figTc}
\end{figure}

The critical temperature versus the disorder strength $\Delta$ calculated in the
mean field approximation is shown in Fig. \ref{figTc}. The critical temperature grows
strongly with disorder. In contrast, the evaluation of the critical temperature
from the spin wave stiffness (\ref{TcSw}) gives a much weaker dependence
on the disorder.

\begin{figure}
\psfig{file=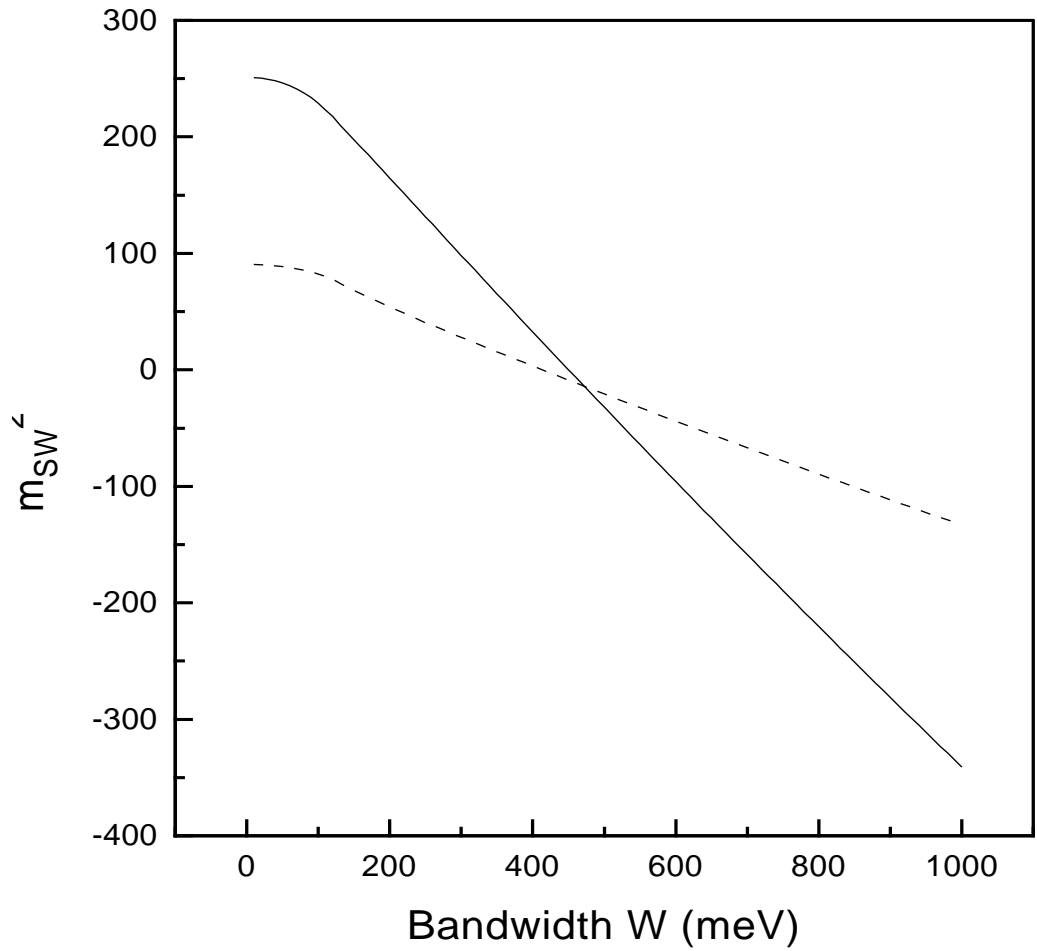,width=15cm,height=20cm,angle=0}
\vskip -3cm
\caption{Zero temperature spin wave mass versus the bandwidth $W$.
$J=10$ meV nm$^3$, $N_p=5, \mu=0$, $\rho=1.0$ meV$^{-1}$ nm$^{-3}$. Solid line: $\Delta=0.5$.
Dashed line: $\Delta=0.3$.}
\label{figmsw}
\end{figure}

The zero temperature mass of the spin wave excitations $m_{SW}$ is given by
\begin{equation}
m^2_{SW}\approx\frac{64\Delta^2 J^2S}{N_p^2}\pi\rho
(\Lambda^0_{\uparrow\uparrow}-\Lambda^0_{\downarrow\downarrow})
-\frac{JS}{2\pi\Delta^2}\int d\omega\left\{(\Lambda^\omega_{\uparrow\uparrow})^2+
(\Lambda^\omega_{\downarrow\downarrow})^2\right\},
\label{SWmass}
\end{equation}
where we neglected a small imaginary part.
The dependence of the spin wave mass $m_{SW}^2$ on the hole bandwidth $W$
at two different sets of other parameters is shown in Fig. \ref{figmsw}.
At large bandwidth  $W$,  $m_{SW}^2$ becomes negative
(see Fig. \ref{figmsw}),
which signals the breakdown of the collinear ferromagnetic order.
The transition to
negative $m_{SW}^2$ occurs at lower values of $W$ for a smaller average number of spins $N_p$
in a magnetic cluster.

Presumably, in the regime $m_{SW}^2<0$ the ground state is noncollinear.
The existence of a
noncollinear ground state in disordered (III,Mn)V semiconductors has been suggested
recently \cite{Schliemann}.

\section{Summary and discussion}

In this paper we investigated the influence of disorder on the critical temperature
of an itinerant ferromagnet. The theoretical modeling in the regime close to the
phase transition was based on the picture of
magnetic polaronic clouds (magnetic clusters) with large relaxation time.
The clusters are
polarized by mobile holes hopping between them.
The regime of small concentration of mobile holes
considered here is relevant for the ferromagnetism in (III,Mn)V compounds.
The disorder in the Mn concentration is
naturally included as a random number of spins in each magnetic cluster.
At the critical point, the clusters begin to percolate, developing
infinite range magnetic correlations.

We found that the critical temperature grows with disorder,
both in the mean field approximation
and if calculated from the zero temperature spin stiffness.
However, the influence of disorder
turns out to be much stronger in the mean field evaluation of $T_c$ than by the estimation
from the spin stiffness. In both approaches, disorder is
taken into account perturbatively. In the mean field calculation, 
the perturbative treatment of disorder can be justified, if
the critical temperature is high (which is the case in the experiment), and the
localization is suppressed.

In contrast, the estimation of $T_c$ from the zero
temperature spin stiffness is much less reliable. 
At zero temperature, the higher order localization corrections become important, and  
perturbative account for disorder may result in the underestimation of the disorder 
effect on the spin stiffness. The latter, in turn, leads to the underestimaiton of 
the influence of disorder on $T_c$.   
At the same time, the calculations of the spin stiffness correctly reflect the general
tendency that the spin stiffness, and therefore  the critical temperature,
grows with disorder.

The physical reason for the increase of $T_c$ at weak disorder was
proposed in Ref. \cite{Bhatt}.
In the disordered system, a mobile hole spends more time in the regions with higher Mn
concentration, thus the effective magnetic coupling increases.
However, at strong disorder, the critical temperature should eventually decrease
because of localization
of the charge carriers that mediate the magnetic correlations.

In this work we neglected the localization corrections that should make the motion
of the holes diffusive. The work on this subject is in progress.

We thank  A. H. MacDonald, J. Schliemann, J. K\"onig, and
K. Scharnberg for useful discussions.

\newpage

\end{document}